\documentclass[aps,prl,twocolumn,superscriptaddress,reprint,groupedaddress]{revtex4}

\usepackage{graphicx}
\usepackage{dcolumn}
\usepackage{bm}
\usepackage{epstopdf}
\usepackage{xr}
\usepackage{amsmath}
\usepackage{color}
\usepackage{xcolor}
\usepackage[normalem]{ulem}
\usepackage{wasysym}
\usepackage{siunitx}
\usepackage{physics}
\usepackage{lineno}
\usepackage{gensymb}
\usepackage[T1]{fontenc}
\usepackage{booktabs}
\usepackage[section]{placeins}

\usepackage{textcomp}
\usepackage{afterpage}
\usepackage[colorlinks=true, linkcolor=black, urlcolor=blue, citecolor=blue]{hyperref}
\usepackage{blindtext}

\makeatletter

\newcommand{\Rmnum}[1]{\expandafter\@slowromancap\romannumeral #1@}
\newcommand{\moire}{moir\'e }
\makeatother

\definecolor{mygreen1}{rgb}{0, 0.5, 0.7}
\graphicspath{ ../Fig/ }

\usepackage{eso-pic}
\usepackage{lipsum}
\usepackage{fancyhdr}
\fancypagestyle{plain}{%
	\fancyhead[R]{\fbox{to appear in journal xx}}
	
}

\begin{document}
	
\title{Tunable bandwidths and gaps in twisted double bilayer graphene system on the verge of correlations}

\author{Pratap Chandra Adak}
\thanks{These two authors contributed equally}
\affiliation{Department of Condensed Matter Physics and Materials Science, Tata Institute of Fundamental Research, Homi Bhabha Road, Mumbai 400005, India.}
\author{Subhajit Sinha}
\thanks{These two authors contributed equally}
\affiliation{Department of Condensed Matter Physics and Materials Science, Tata Institute of Fundamental Research, Homi Bhabha Road, Mumbai 400005, India.}
\author{Unmesh Ghorai}
\affiliation{Department of Theoretical Physics, Tata Institute of Fundamental Research, Homi Bhabha Road, Mumbai 400005, India.}
\author{L. D. Varma Sangani}
\affiliation{Department of Condensed Matter Physics and Materials Science, Tata Institute of Fundamental Research, Homi Bhabha Road, Mumbai 400005, India.}
\author{Kenji Watanabe}
\affiliation{National Institute for Materials Science, 1-1 Namiki, Tsukuba 305-0044, Japan.}
\author{Takashi Taniguchi}
\affiliation{National Institute for Materials Science, 1-1 Namiki, Tsukuba 305-0044, Japan.}
\author{Rajdeep Sensarma}
\homepage{sensarma@theory.tifr.res.in}
\affiliation{Department of Theoretical Physics, Tata Institute of Fundamental Research, Homi Bhabha Road, Mumbai 400005, India.}
\author{Mandar M. Deshmukh}
\homepage{deshmukh@tifr.res.in}
\affiliation{Department of Condensed Matter Physics and Materials Science, Tata Institute of Fundamental Research, Homi Bhabha Road, Mumbai 400005, India.}

\begin{abstract}
	We use temperature-dependent resistivity in small-angle twisted double bilayer graphene to measure bandwidths and gaps of the bands. This electron-hole asymmetric system has one set of
	non-dispersing bands that splits into two flat bands with the electric field - distinct from the twisted bilayer system. With electric field, the gap between two emergent flat bands increases monotonically and bandwidth is tuned from 1 meV to 15 meV. These two flat bands with gap result in a series of thermally induced insulator to metal transitions - we use a
	model, at charge neutrality, to measure the bandwidth using only transport measurements. Having two flat bands with tunable gap and bandwidth offers an opportunity to probe the emergence of correlations.
\end{abstract}

\maketitle

 The ability to tune the twist angle between two sheets of two-dimensional materials
 has enormously increased the scope of engineering superlattices~\cite{li_observation_2010,wang_fractal_2012,cao_superlattice-induced_2016}. Following theoretical predictions~\cite{bistritzer_moire_2011,suarez_morell_flat_2010,lopes_dos_santos_continuum_2012}, some recent experiments have confirmed flat bands for twisted  bilayer graphene system near a magic angle. In addition, for some parameter space, it hosts highly correlated states showing exotic phenomena including Mott insulator, superconductivity, and ferromagnetism~\cite{cao_correlated_2018,cao_unconventional_2018,kim_tunable_2017-2,sharpe_emergent_2019-1,lu_superconductors_2019}. The ratio of electron-electron interaction strength and width of the electronic band is a key parameter determining the emergence of correlations. It is desirable to find additional knobs to tune the bandwidth~\cite{yankowitz_tuning_2019,chen_evidence_2019,chen_signatures_2019}. 
 
 \begin{figure}[hbt]
 	\includegraphics[width=8.6cm]{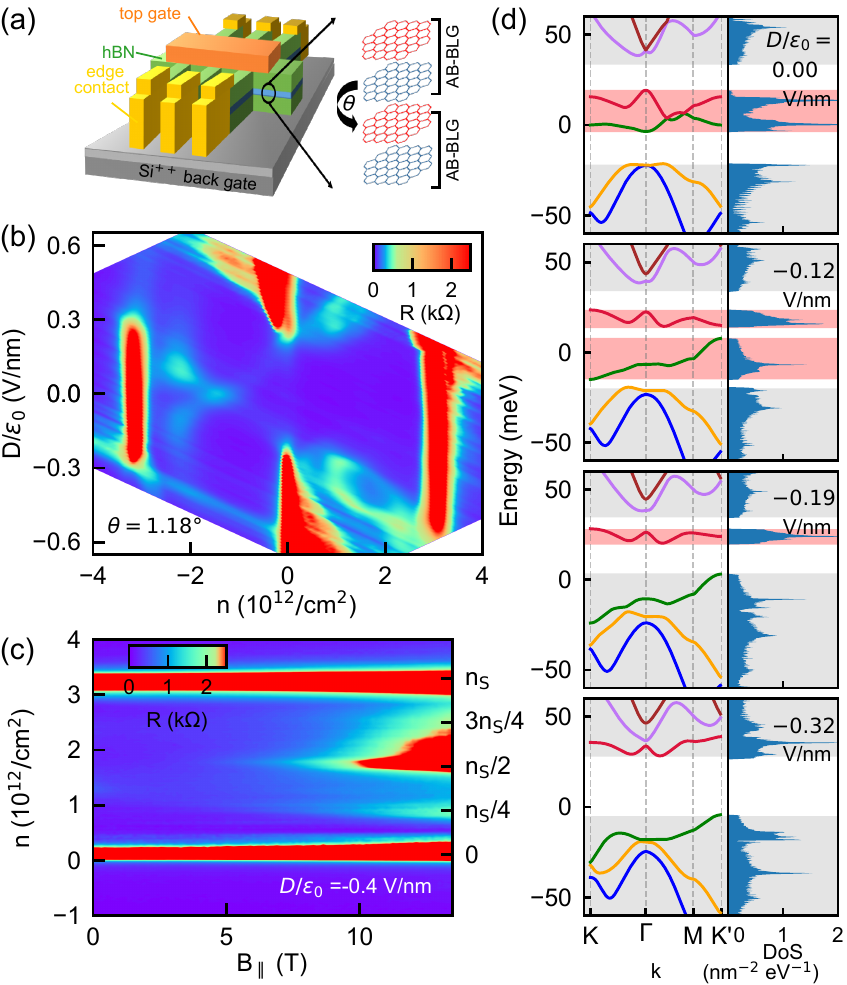}
 	\caption{ \label{fig:fig1} {\footnotesize \textbf{The TDBG device, resistance variation and DoS distribution.} (a) The schematic of the dual gated twisted double bilayer graphene device. (b) Two-dimensional color scale plot of four-probe resistance as a function of carrier density ($n$) and the perpendicular electric displacement field ($D$). Gaps are  formed for $n=0$ and $\pm n_\text{S}$.  (c) Development of correlation induced gaps at $D/\epsilon_0=-0.4$~V/nm with the in-plane magnetic field up to 13.5 T. (d) The calculated low energy band structure (left panel) and corresponding density of states (right panel) at $B_{\parallel}=0$ illustrating the sequence of opening and closing of different gaps qualitatively. The energy range of the flat bands and the dispersing bands are shaded in pink and grey, respectively. }}
 \end{figure}

\afterpage{\clearpage}

 Twisted double bilayer graphene (TDBG) is also an interesting platform where one puts a copy of bilayer graphene on top of another one with a twist angle between them. The Bernal-stacked bilayer graphene has a unique band structure that changes  upon applying perpendicular electric field as a bandgap opens up with increasing field~\cite{castro_biased_2007,oostinga_gate-induced_2008,zhang_direct_2009,castro_neto_electronic_2009-1}. The TDBG system inherits this tunability with the electric field and makes itself an interesting system with electrically tunable flat bands~\cite{chebrolu_flat_2019,zhang_nearly_2019,koshino_band_2019-1,lee_theory_2019,choi_intrinsic_2019}. However, there are only few experimental reports so far on this novel system  where electric field tunable correlated insulating states and the realization of superconductivity have been demonstrated~\cite{shen_observation_2019,liu_spin-polarized_2019,cao_electric_2019,burg_correlated_2019}. There is an urgent need to understand the single-particle band structure and this requires a detailed study of electron transport in small-angle TDBG system to understand bandgaps, bandwidths, and their tunability with the electric field.

 In this work, we present a comprehensive study of temperature dependence of electronic transport in a twisted double bilayer graphene device with a twist angle of $1.18\degree$ between the two sheets of bilayer graphene. This electron-hole asymmetric system has one set of low-energy non-dispersing bands separated from higher energy dispersing bands by \moire gaps at zero perpendicular electric field. As we increase electric field more, two clear trends emerge -- firstly, after a finite electric field a bandgap opens up at the charge neutrality point giving rise to two flat bands; this differentiates it from twisted bilayer graphene, where the low energy bands always touch at the Dirac points~\cite{chebrolu_flat_2019,koshino_band_2019-1}. Secondly, further increase of electric field closes the \moire gaps gradually in an asymmetric manner for electron and hole side. By systematic measurement of gaps as a function of electric field we establish both these trends experimentally. We then focus on understanding the thermally induced insulator to metal transitions in the parameter space of electric field and charge density. We show that this physics can be understood using a simple model with two flat bands at a finite perpendicular electric field. Applying this simple double quantum well with flat band model~\cite{mandrus_thermodynamics_1995} we extract, using only transport measurements, the bandwidth tunable from $\sim 1$~meV to $15$~meV by increasing the electric displacement field strength, $D/\epsilon_0$ from $-0.24$~V/nm to $-0.34$~V/nm. The \moire  gaps separating the flat bands from the dispersing higher bands are $\sim 10$~meV. The fact that the gaps and bandwidths are tunable with the electric field makes this system important for studying correlations.

We now discuss the details of our experiment. As shown in the schematic in Fig.~\ref{fig:fig1}(a), the twisted double bilayer graphene is encapsulated by both top and bottom few layers thick hexagonal boron nitride~\cite{wang_one-dimensional_2013-1,kim_van_2016,cao_superlattice-induced_2016}. For details of the fabrication process, see Section I in Supplemental Material~\cite{suppl}. The metallic top gate and the Si$^{++}$ back gate allow us independent control of both the carrier density $(n)$ and the perpendicular electric displacement field $(D)$, given by the equations $n=(C_{TG}V_{TG}+C_{BG}V_{BG})/e$ and $D=(C_{TG}V_{TG}-C_{BG}V_{BG})/2$, where $C_{TG}$ and $C_{BG}$ are the capacitance per unit area of the top and the back gate respectively, $e$ being the charge of an electron.

Figure \ref{fig:fig1}(b) shows a color scale plot of the four-probe resistance, $R$, as a function of $n$ and $D$ measured at a temperature of 1.5~K.  Firstly, this TDBG device shows two pronounced peaks in resistance corresponding to the two \moire gaps at $n=\pm n_\text{S}$, where $n_\text{S}=3.2 \times 10^{12}$~cm$^{-2}$ is the required charge density to fill the first \moire band corresponding to a twist angle of $1.18\degree$. Another peak in resistance appears at the charge neutrality point (CNP) as the magnitude of $D$ is increased indicating gap opening with the electric field. This is also seen in simple bilayer graphene device~\cite{oostinga_gate-induced_2008,zhang_direct_2009}. However, the resistance at the two \moire gaps decreases with increasing electric field, indicating gradual closing of these gaps.  Secondly, as reported in some very recent works on TDBG~\cite{shen_observation_2019,liu_spin-polarized_2019,cao_electric_2019,burg_correlated_2019} there are several interesting features at specific parameter space, namely, a cross-like feature around zero electric field only on the hole side and two halo regions of high resistance  present at the electric displacement field $D/\epsilon_0 \sim \pm 0.4$ V/nm on the electron side. In this device, we do not observe fully formed insulating states at filling other than $n=\pm n_\text{S}$ and $n=0$ at zero magnetic field. We intentionally use this system as we do not want any of the gaps to be enhanced due to correlations. However, Fig.~\ref{fig:fig1}(c) shows that with the in-plane magnetic field, gaps open up at $n_\text{S}/4$, $n_\text{S}/2$ and $3n_\text{S}/4$ for this device -- suggesting that the system is on the verge of developing correlations. All of these observations indicate the existence of flat band physics and offer an opportunity to study the temperature dependence of transport properties in a flat band system. See Fig.~S8 in Supplemental Material~\cite{suppl} for data from another TDBG device with a twist angle $2.05\degree$ that shows two \moire gaps at $n=\pm n_\text{S}$ and gap opening at CNP, but no other features like the cross or the halo. This emphasizes the role of the flat bands in the device with $1.18\degree$ twist angle.

\begin{figure}[hbt]
	\includegraphics[width=8.6cm]{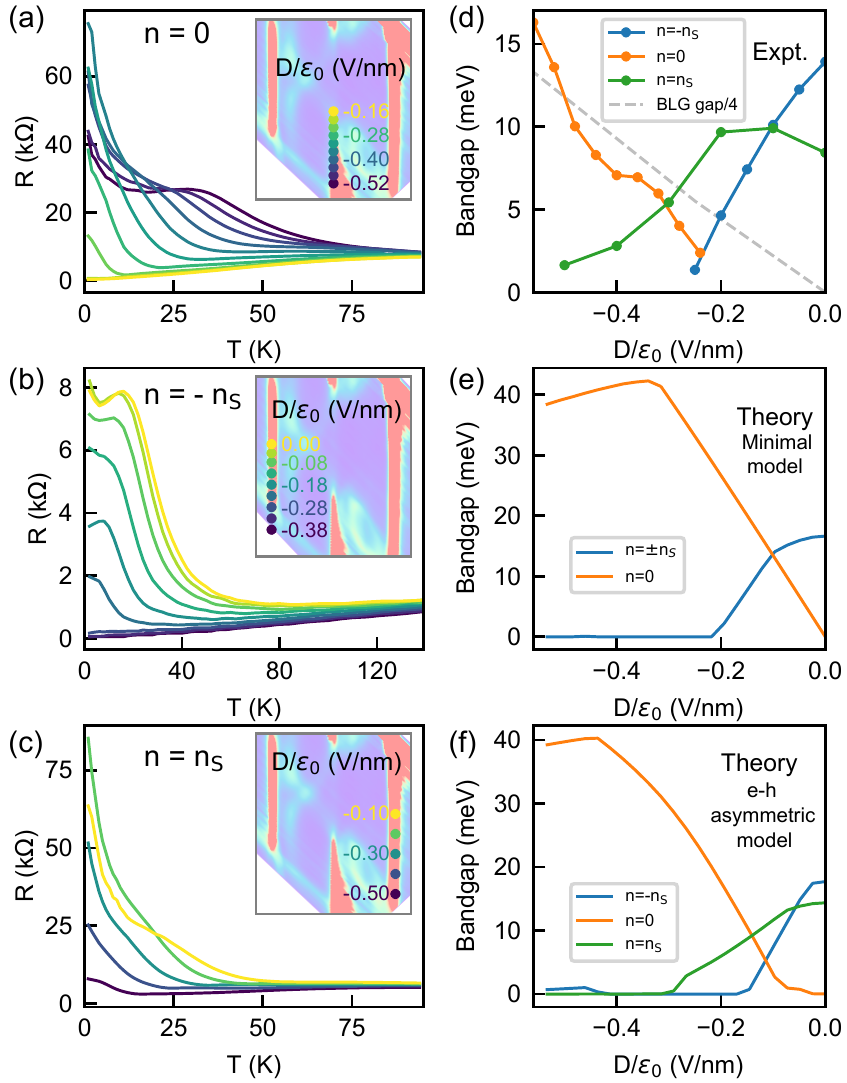}
	\caption{ \label{fig:fig2} {\footnotesize \textbf{Temperature variation of resistance at various insulating gaps and estimation of Arrhenius gap.}  (a-c) Line slices of resistance vs. temperature for different electric fields at $n=0$, $n=-n_\text{S}$, and $n=n_\text{S}$ respectively. Insets show the locations of the data points in $(n,D)$ space. (d) Variation of bandgap extracted by Arrhenius fittings to the temperature variation of resistance for $n=0$ and $n=\pm n_\text{S}$. The gap of bilayer graphene, as in Y. Zhang et al.~\cite{zhang_direct_2009}, is plotted with dashed grey line to show the dissimilarity between CNP gap in BLG and TDBG. (e-f) Variation of the bandgap calculated using minimal model (e) and electron-hole asymmetric model (f), respectively; gap opening at CNP only after a critical field seen in experiments is captured only by the electron-hole asymmetric model.}}
\end{figure}

From Fig.~\ref{fig:fig1}(b) it is evident that this system possesses a high degree of electron-hole asymmetry. While the asymmetric presence of above mentioned cross-like feature and the halo feature are such examples, the different closing sequence of two \moire gaps offers another one.  Different theoretical aspects of twisted angle systems have been understood using a minimal model where one sets  further neighbor hoppings across the bilayer, $\gamma_3$ and $\gamma_4$, as well as the potential on dimer sites $\Delta'$ to be zero. However, to capture the electron-hole asymmetry we use an e-h asymmetric model with  $\gamma_3 = 320$ meV, $\gamma_4 = 44$ meV, and $\Delta' = 50$ meV (see Section II in Supplemental Material~\cite{suppl} for details). In Fig.~\ref{fig:fig1}(d) (left panels) we calculate the band structures of TDBG for a twist angle of $1.18\degree$ for four different values of electric fields based on this e-h asymmetric model. The corresponding density of states is plotted in the right panels of  Fig.~\ref{fig:fig1}(d), which depicts various gaps in agreement with those seen from the resistance variation in Fig.~\ref{fig:fig1}(b). 

\begin{figure*}
	\includegraphics[width=15cm]{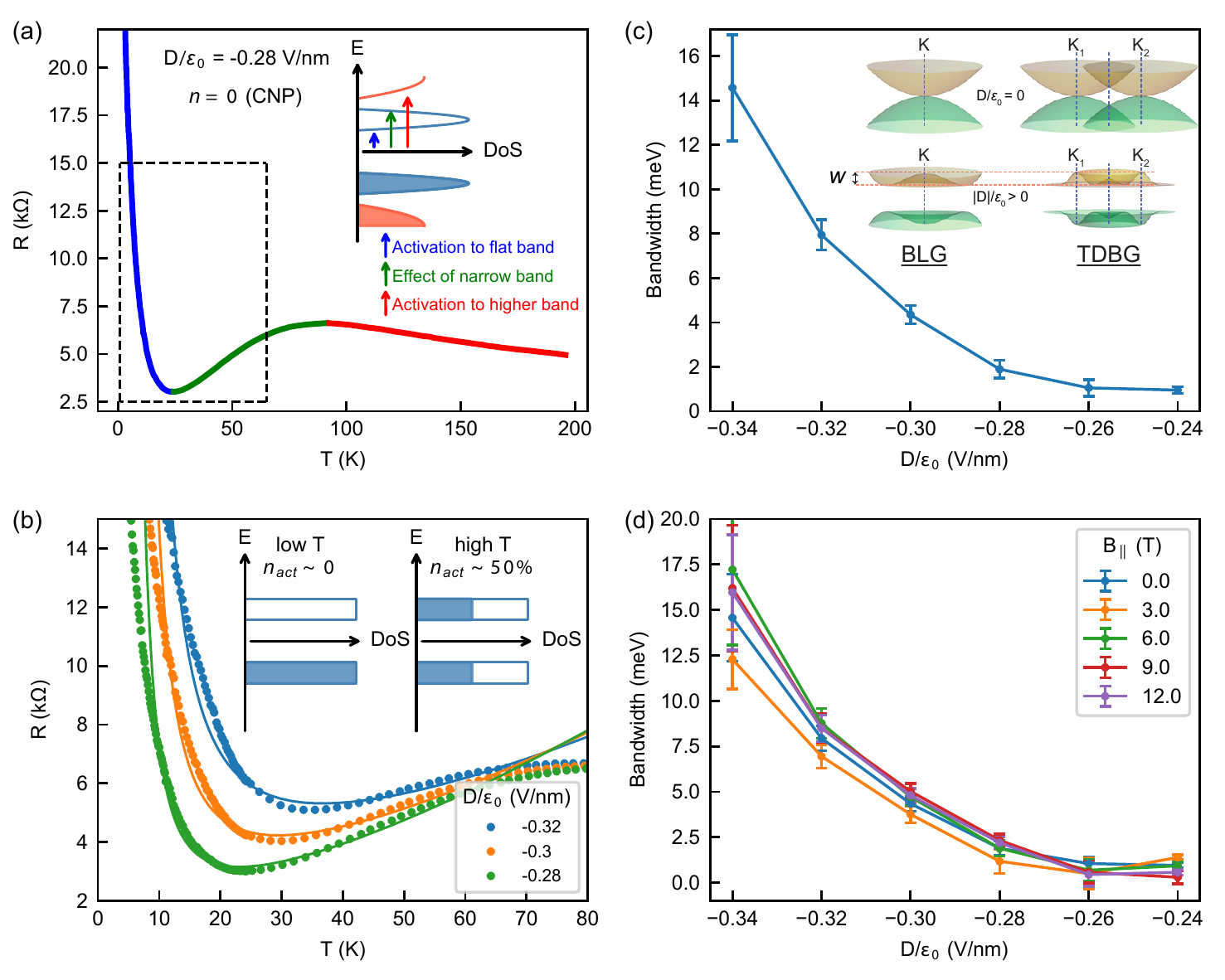}
	\caption{ \label{fig:fig3} {\footnotesize \textbf{Insulator to metal crossover - extracting bandwidth.} (a) A representative $R$ vs. $T$ curve for $n=0$ depicting three regions of different temperature dependence -- insulating behaviour at low $T$ (blue), metallic increase (green), and insulating behaviour again (red). Inset depicts the three corresponding energy scales in the simple schematic of the underlying band structure with CNP gap and two \moire gaps. The dashed black rectangle indicates the portion of the curve used in (b). (b) Fitting insulator to metal crossover in $R$ vs. $T$ curves for different electric fields at $n=0$ to extract bandwidth. Points correspond to measured resistance and the solid lines are the fits. Inset shows the simple model of two flat bands with a small gap used in the extraction. The number of activated charge carriers saturates to 50\% at higher temperature from its low $T$ value of $n_{act}\sim 0$. (c) The variation of extracted bandwidth~($w$) with the electric field at $B_{\parallel}=0$. The inset shows how TDBG can inherit the tunability of its bandwidth from BLG. (d) The variation of bandwidth~($w$) with the electric field for different in-plane magnetic fields. }}
\end{figure*}

 We now probe the electron-hole asymmetry in detail by extracting the values of different gaps with varying electric field. Figures \ref{fig:fig2}(a-c) show the variation of resistance as a function of temperature for different electric fields at $n=0$, $n=-n_\text{S}$, and $n=n_\text{S}$ respectively. We extract gaps, $\Delta$, by fitting $R\propto \exp(\Delta/2k_\text{B}T)$ at the regime  where the resistance decreases exponentially with the temperature. See Section III in Supplemental Material~\cite{suppl} for details of the fits. The $R$ vs. $T$ curves show deviation from the Arrhenius nature at low temperature. In particular, we see a shoulder like variation, which is most prominent for $n=0$ with $D/\epsilon_0$ stronger than $-0.36$~V/nm (see Section IV in Supplemental Material~\cite{suppl} for details). In  Fig.~\ref{fig:fig2}(d) we plot extracted gaps as a function of the perpendicular electric field. Consistent with the band picture we presented in the earlier section, the gap at the charge neutrality point increases with the electric field, while the gaps for $n=\pm n_\text{S}$ decrease. An interesting observation is that the gap at CNP  starts opening only after a finite electric displacement field of $D/\epsilon_0 \sim 0.2$ V/nm, in contrast to bilayer graphene where the gap opens up as soon as any non-zero electric field is applied (see Fig.~S9 in Supplemental Material~\cite{suppl}). A similar analysis for the device with a twist angle $2.05\degree$ shows  non-monotonic behavior of gap with the electric field as the CNP gap closes and reopens with the increasing electric field (see  Fig.~S8 in Supplemental Material~\cite{suppl}). In Fig.~\ref{fig:fig2}(e) and \ref{fig:fig2}(f), we present the calculated gaps based on the minimal model and the e-h asymmetric model respectively. The opening of the CNP gap only after $D/\epsilon_0 \sim 0.2$ V/nm shows the inadequacy of the minimal model, which indicates a finite gap at any non zero electric field. However, the e-h asymmetric model captures the correct trend of the variation of gaps with the electric field for all three different values of $n=0,\pm n_\text{S}$, though overestimating the gaps quantitatively.

In Fig.~\ref{fig:fig3}(a) we plot a typical resistance vs. temperature curve at $n=0$ for $D/\epsilon_0 = -0.28$~V/nm when the Fermi energy lies in the CNP gap between two flat bands separated from higher band by \moire gaps as shown in the inset of Fig.~\ref{fig:fig3}(a). Clearly, there are three distinct temperature regions governed by three different energy scales -- the CNP gap, the flat band width and the separation of higher-energy bands from the Fermi energy. While resistance shows insulating behavior at low temperature (blue) due to activation of carriers to flat band, at high temperature (red) this is governed by the activation to higher dispersing band. The thermal activation of charge carriers to higher dispersing bands is also captured by the metal to insulator transition for the states where the chemical potential lies within the flat band, not just at CNP (see Section VII in Supplemental Material~\cite{suppl} for details)~\cite{polshyn_large_2019}.

However, there is a mid-range temperature regime (green) where resistance increases with temperature. We now closely examine the temperature regime within the dashed box in Fig.~\ref{fig:fig3}(a) and illustrate how the existence of two flat bands separated by a narrow gap can account for this crossover. We employ a simple model, as depicted in the inset of Fig.~\ref{fig:fig3}(b) where we consider two narrow bands with bandwidth of $w$ having the constant density of states $\text{DoS}(E) =n_\text{S}/w$ with a narrow gap of $\Delta$ between them. Here $n_\text{S}$ is the number of electrons to  fill one such band. We note, this model is most appropriately valid before the closure of the \moire gap. As the temperature is increased, the carriers are activated in the narrow bands governed by the Fermi-Dirac distribution, which results in the usual decrease in resistance. However, due to the small energy scales set by $w$ and $\Delta$, the number of activated carriers  quickly saturates to the value one would get at a very high temperature, $k_\text{B}T>>\Delta, w$ (see Section VI in Supplemental Material~\cite{suppl}). On the other hand,  the mobility keeps on decreasing with increasing temperature and takes over the increase in activated carrier density to show a metal-like increase in resistance. 

To quantitatively understand this insulator to metal transition seen in Fig.~\ref{fig:fig3}(b), we assume a phonon-scattering dominated temperature variation of the mobility of the form $\mu=a\times T^{-3/2}$. We calculate the conductance using the relation $\sigma=n e \mu$. Here $e$ is the charge of electron and $n$ is the number of activated carriers given by $n=\int \text{DoS}(E) f(E,E_F,T) dE$ with $f(E,E_F,T)$ being the Fermi-Dirac function. We then fit the resultant expression of the conductivity with three fitting parameters, namely, the bandwidth $(w)$, the bandgap $(\Delta)$, and the overall factor $a$, to the experimental data. Few such fitting curves are overlayed with the experimental data in Fig.~\ref{fig:fig3}(b). While it deviates at high temperature due to carrier activation to the higher band, the low temperature deviation can be understood since the simple picture of the  mobility decreasing with temperature does not hold good.

 In Fig.~\ref{fig:fig3}(c), we plot the extracted bandwidths as a function of the electric field. The bandwidth has a clear trend of increasing with the electric field. This reasonably agrees with the results discussed in N. R.  Chebrolu et al.~\cite{chebrolu_flat_2019}. The bandwidth has a typical value $\sim 1-15$ meV for mid-range of the electric field where the halo features appear on the electron side. 
 
 The origin of the electric field tunability of the bandwidth is subject of detail theoretical and numerical calculation~\cite{chebrolu_flat_2019}. However, a simple picture is depicted in the inset of Fig.~\ref{fig:fig3}(c) considering the tunability of underlying BLG band structure. Upon applying electric field low energy part of quadratic bands in BLG forms a ``Mexican hat" structure with local maxima and minima around $K$ points in the Brillouin zone~\cite{castro_biased_2007,castro_neto_electronic_2009-1}. The depth of the Mexican hat scales approximately as the cube of the electric field. Small angle twist of  two copies of BLG in TDBG effectively hybridizes such adjacent $E-k$ cones and the bandwidth of the low energy flat band inherits the electric field dependence of the depth of the Mexican hat in BLG. The tunability of this energy scale may be enhanced in TDBG from that of BLG due to additional hopping terms and modified effective mass. 
 
 Finally, we study the effect of in-plane magnetic field on the bandwidth, motivated by its profound role to enhance correlated gaps as already seen in Fig.~\ref{fig:fig1}(c). In Fig.~\ref{fig:fig3}(d) we plot the variation of bandwidth with electric field for different magnitude of in-plane magnetic field. The overlapping error bars suggest there is no clear trend implying that the bandwidth is not substantially tuned with in-plane magnetic field.

 We find TDBG to be a highly tunable system with \moire gaps that can be closed with increasing electric field while the flat bands split into two; eventually, the two narrow flat bands merge with the higher dispersing bands. A plethora of novel phenomena have been identified to be originating from the underlying flat bands in twisted graphene system, without directly quantifying the flatness of the bands using transport. In strongly correlated systems, the bandwidths and gaps can be enhanced because of interactions. Our choice of a weakly interacting system and subsequent analysis provide a route to quantitatively establish tunability of bandwidth with electric field in flat band systems for the first time.

We thank Allan MacDonald, Senthil Todadri, Sreejith GJ, Priyanka Mohan and Biswajit Datta for helpful discussions. We acknowledge Nanomission grant SR/NM/NS-45/2016 and Department of Atomic Energy of Government of India for support.  U. G. and R. S. thanks the computational facility at Department of  Theoretical Physics, TIFR, Mumbai. Preparation of hBN single crystals is supported by the Elemental Strategy Initiative conducted by the MEXT, Japan and JSPS KAKENHI Grant Number JP15K21722.

%

\end{document}


\newcommand {\beq} {\begin{equation}}
\newcommand {\eeq} {\end{equation}}
\newcommand {\bqa} {\begin{eqnarray}}
\newcommand {\eqa} {\end{eqnarray}}

\newcommand{\kbf}{\textbf{k}}
\newcommand{\Kbf}{\textbf{K}}
\newcommand{\abf}{\textbf{a}}
\newcommand{\bbf}{\textbf{b}}
\newcommand{\Gbf}{\textbf{G}}


\begin{textblock}{5}(6,1)
	\noindent\Large Supplemental Material
\end{textblock}
\title{Tunable bandwidths and gaps in twisted double bilayer graphene system on the verge of correlations}

\author{Pratap Chandra Adak}
\thanks{These two authors contributed equally}
\affiliation{Department of Condensed Matter Physics and Materials Science, Tata Institute of Fundamental Research, Homi Bhabha Road, Mumbai 400005, India}
\author{Subhajit Sinha}
\thanks{These two authors contributed equally}
\affiliation{Department of Condensed Matter Physics and Materials Science, Tata Institute of Fundamental Research, Homi Bhabha Road, Mumbai 400005, India}
\author{Unmesh Ghorai}
\affiliation{Department of Theoretical Physics, Tata Institute of Fundamental Research, Homi Bhabha Road, Mumbai 400005, India}
\author{L. D. Varma Sangani}
\affiliation{Department of Condensed Matter Physics and Materials Science, Tata Institute of Fundamental Research, Homi Bhabha Road, Mumbai 400005, India}
\author{Kenji Watanabe}
\affiliation{National Institute for Materials Science, 1-1 Namiki, Tsukuba 305-0044, Japan}
\author{Takashi Taniguchi}
\affiliation{National Institute for Materials Science, 1-1 Namiki, Tsukuba 305-0044, Japan}
\author{Rajdeep Sensarma}
\homepage{sensarma@theory.tifr.res.in}
\affiliation{Department of Theoretical Physics, Tata Institute of Fundamental Research, Homi Bhabha Road, Mumbai 400005, India}
\author{Mandar M. Deshmukh}
\homepage{deshmukh@tifr.res.in}
\affiliation{Department of Condensed Matter Physics and Materials Science, Tata Institute of Fundamental Research, Homi Bhabha Road, Mumbai 400005, India}

%
%
\maketitle



\section{SAMPLE PREPARATION and measurement technique}
The devices were fabricated using the standard poly(propylene) carbonate (PPC) based dry transfer method for stacking~\cite{wang_one-dimensional_2013-1}. For making  the hBN-Gr-Gr-hBN stacked device the individual flakes were singled out optically. The hBN flakes are chosen  to be between 20~nm to 40~nm in thickness, measured by AFM. The number of Graphene layers was determined by optical contrast. Subsequently, it was cut into two halves. After picking up one of the cut pieces with hBN, the other piece left on the bottom stage was rotated slightly more than the desired angle and picked, followed by another hBN pickup. The resulting stack was dropped on $\text{SiO}_2$/Si substrate and processed without thermal annealing. For the 2.05$^\circ$ device, the two separate bilayer graphenes were aligned visually to make an hBN-Gr-Gr-hBN stack, and dropped on a graphite flake to make a device with bottom graphite-gate. The geometry of the devices was defined by e-beam lithography and shaped via etching in $\text{CHF}_3$+$\text{O}_2$ plasma. The top gate and electrical contacts were deposited by e-beam evaporation of Cr/Au and Cr/Pd/Au, respectively.

The transport measurements were carried out in a He-4 flow cryostat using low frequency ($\sim$13-17 Hz) lock-in method. A current of $\sim$10~nA was sent, and the four-probe voltage was measured using lock-in amplifier after amplifying with a preamplifier. The twist angle ($\theta$) was calculated from $n_\text{S}$ where $n_\text{S}=8\theta^{2}/(\sqrt{3}a^{2})$, $a=0.246$~nm is the lattice constant of graphene. The temperature was measured using a diode sensor~(DT-470) placed very close to the sample.

\clearpage  
\section{Band structure calculation}
\begin{figure}
	\centering
	\includegraphics[width=15cm]{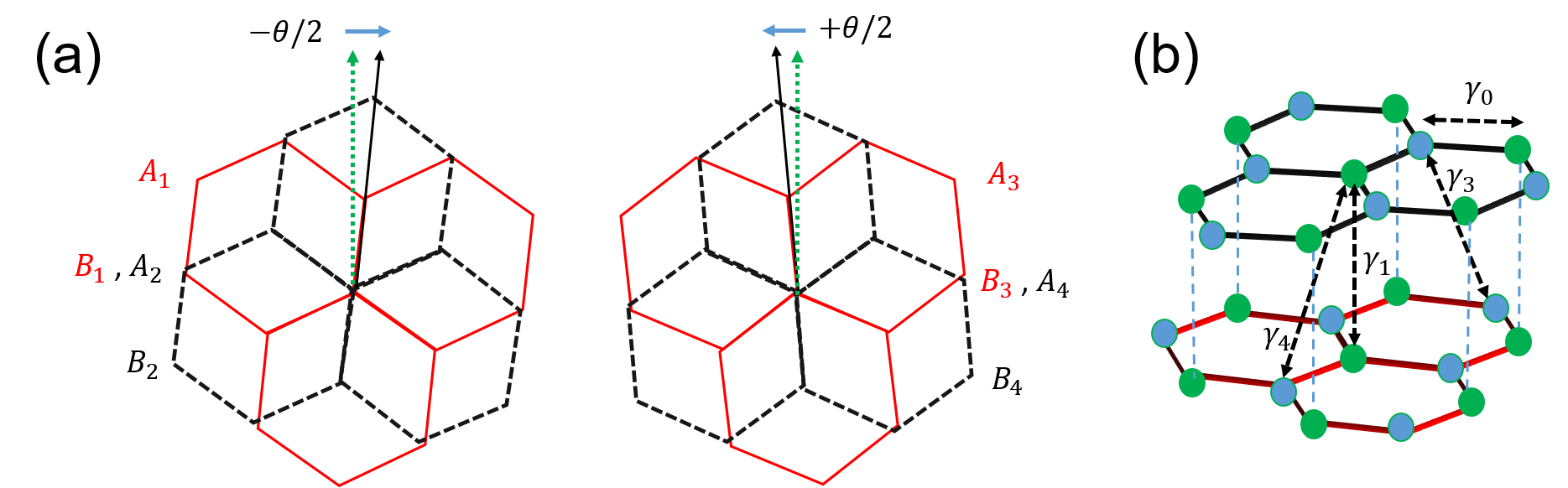}
	\caption{ \label{fig:fig_theory}  (a) Atomic structure of twisted AB-AB double bilayer. (b) Inter-layer couplings between the graphene bilayer.}
\end{figure}
We consider an AB-AB twisted double bilayer graphene where each layer is a Bernal-stacked BLG. At zero twist angle, the stacking of the layers is given by A-B-A-B. We consider AB bilayers twisted by an angle $\theta$ (see Fig.~\ref{fig:fig_theory}(a)). 
We describe our system by an effective continuum Hamiltonian constructed in the basis of Bloch wavefunctions of the sublattices $(A_1,B_1,A_2,B_2,A_3,B_3,A_4,B_4)$ as ~\cite{koshino},\\
\begin{equation}
	H_{AB\text{-}AB}=\begin{pmatrix}
		H_0(\kbf_1)&g^\dagger(\kbf_1)&0&0\\g(\kbf_1)&H^{'}_{0}(\kbf_1)&U^\dagger&0\\0&U&H_0(\kbf_2)&g^\dagger(\kbf_2)\\0&0&g(\kbf_2)&H^{'}_{0}(\kbf_2)
	\end{pmatrix}+V
\end{equation}  
where, $\kbf_l=R(\pm\theta/2)(\kbf-\Kbf^{(l)}_{\xi})$. Here, $R(\theta)$ is the matrix for 2-D rotation by angle $\theta$ 
and $\pm$ sign is for top ($l$=1) and bottom ($l$=2) layers respectively. $\xi$ denotes the valley index and takes values, $\pm1$. $\Kbf_{\xi}^{l}=R(\mp\theta/2)\Kbf_{\xi}^{D}$ where the rotated Dirac point wave vector $\Kbf_{\xi}^{D}=-\xi(4\pi/3a)(1,0)$, $a\approx0.246$~nm being the graphene lattice constant and\\
\begin{equation}
	H_{0}(\kbf)=\begin{pmatrix}
		0&-\hbar vk_{-}\\-\hbar vk_{+}&\Delta^{'}
	\end{pmatrix},H_{0}^{'}(\kbf)=\begin{pmatrix}
		\Delta^{'}&-\hbar vk_{-}\\-\hbar vk_{+}&0
	\end{pmatrix},g(\kbf)=\begin{pmatrix}
		\hbar v_4k_{+}&\gamma_1\\ \hbar v_3k_{-}&\hbar v_4k_{+}
	\end{pmatrix}.
\end{equation}
Here, $k_{\pm}=\xi k_x\pm i k_y$. $H_0$ and $H_{0}^{'}$ are the Dirac Hamiltonian for single layer graphene. $\Delta^{'}=50$~meV~\cite{koshino} is the potential difference between dimer sites (i.e. sites which have lattice sites just above/below them) and non-dimer sites. The graphene fermi velocity is $\hbar v/a=2.1354$~eV~\cite{koshino}. The interlayer coupling in the AB bilayer is described ($g$ matrix) in terms of three parameters $\gamma_1$, $\gamma_3$ and $\gamma_4$ as shown in Fig.~\ref{fig:fig_theory}(b). Here we use $\gamma_1=0.4$~eV, $\gamma_3=0.32$ eV and $\gamma_4=0.044$ eV~\cite{koshino}. $v_3$ and $v_4$ are related to $\gamma_3$ and $\gamma_4$ by $v_i=(\sqrt{3}/2)\gamma_i a/\hbar(i=3,4)$.
The interlayer coupling between the layers which had twist of angle $\theta$ between them is represented by,
\begin{equation}
	U=\begin{pmatrix}
		u&u'\\u'&u
	\end{pmatrix}+\begin{pmatrix}
		u&u'w^{-\xi}\\u'w^{\xi}&u
	\end{pmatrix}e^{i\xi\Gbf_{1}^{M}.\textbf{r}}+\begin{pmatrix}
		u&u'w^{\xi}\\u'w^{-\xi}&u
	\end{pmatrix}e^{i\xi(\Gbf_{1}^{M}+\Gbf_{2}^{M}).\textbf{r}}
\end{equation}
where $w=e^{2\pi i/3}$ and $u$ \& $u'$ are the interlayer hopping parameters between A to A sublattice and A to B sublattice respectively. We have taken $u=50$~meV \& $u'=85$~meV. Here $\Gbf_{i}^{M}=\bbf_{i}^{*(1)}-\bbf_{i}^{*(2)}(i=1,2)$ are the \moire reciprocal lattice vectors with $\bbf_{i}^{*(l)}=R(\mp \theta/2)\bbf_{i}^{*}$, where $\bbf_{1}^{*}=(2\pi/a)(1,-1/\sqrt{3})$ and $\bbf_{2}^{*}=(2\pi/a)(0,2/\sqrt{3})$ are the reciprocal lattice vectors of Bilayer Graphene without rotation~\cite{koshino}.  The \moire undulations couple momenta differing by $n\Gbf_{i}^{M}$ where $n$ is an integer. To cover the full BZ of the underlying graphene one needs to consider a large number of such momenta (i.e. large number of extended \moire BZs). We truncate this construction after a small number of extended zones which leads to a 124 dimensional Hamiltonian matrix in our case. The truncation is done by checking the convergence of the low energy dispersion~\cite{bistritzer_moire_2011}.
Finally, V represents the potential difference between layers due to the constant perpendicular electric field applied to the system,\\
\begin{equation}
	V=\begin{pmatrix}
		\frac{3}{2}\Delta \hat{1}&0&0&0\\0&\frac{1}{2}\Delta\hat{1}&0&0\\0&0&-\frac{1}{2}\Delta\hat{1}&0\\0&0&0&-\frac{3}{2}\Delta\hat{1}
	\end{pmatrix}.
\end{equation}
Here $\Delta$ is the potential difference between two consecutive layers. The relation between potential, $\Delta$ and the electric displacement field, $D$ is given by, $\Delta=D d/(\epsilon_0 k_\text{hBN})$, where $d=$ distance between two layers of graphene $\approx$ 0.33 nm and ${k_\text{hBN}}$ = dielectric constant of hBN $\approx$ 4. \\

\begin{figure}[hbt]
	\includegraphics[width=10cm]{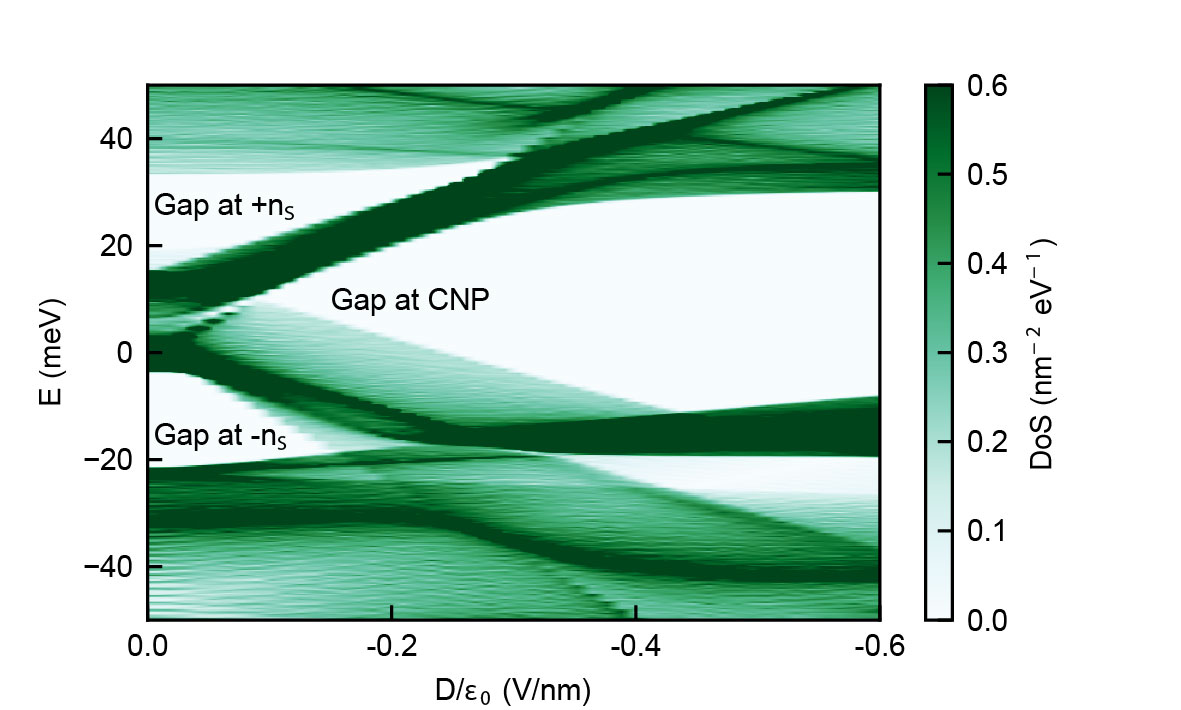}
	\caption{ \label{fig:TheoryDoS} {\footnotesize Color scale plot of calculated DOS as a function of electric field and energy for the $1.18\degree$ TDBG device.}}
\end{figure}

Now in this model, switching off terms $R(\pm\theta/2)$, $\Delta^{'}$, $v_3$ and $v_4$ leads to an electron-hole (e-h) symmetric minimum model. While keeping all these parameters gives an e-h asymmetric full model~\cite{koshino}. Out of these parameters, $v_4$ and $\Delta^{'}$ induces the e-h asymmetry in the system~\cite{asvinspin}. Within the full model, band structure for some of the lowest bands close to zero energy and Density of States (DOS) for few electric field values have been shown in Fig. 1(d) main article. Fig.~\ref{fig:TheoryDoS} shows the evolution of the DOS as a function of electric field. \\

Theoretically, we define bandgap between bands with index $\mu$ and $\mu+1$ as $\Delta_\mu$=Min[$\epsilon_{\mu+1}$] - Max[$\epsilon_{\mu}$] over the whole BZ where $\epsilon$ is the energy of the band. One can see in Fig. 1(d) left panel in the main manuscript, with increasing electric field, the maxima of the flat band shifts from $\Gamma$ point to $K$ point, but the band above still has a minima at $\Gamma$ point which leads to an indirect bandgap. We note that this is a measure of the minimum amount of direct and indirect band gaps. Since \moire BZ corresponds to a length scale $\lambda_\mu>>a$, we believe this is the correct gap to consider for transport experiments. \\

There are three prominent transport gaps in the experiment $n=0, \pm n_\text{S}$ (Fig. 1(b) in the main manuscript). At zero doping (CNP), the chemical potential is close to $\mu=0$. The transport is governed by the presence/absence of low energy states around this energy. The corresponding gap (when it exists) is the splitting between the two flat bands lying closest to $\epsilon=0$ on either side. 
On the other hand, at $n=n_\text{S}$, the upper flat band is filled, and the gap is the splitting between the upper flat band and the next higher energy band (referred as \moire gap). At $n=-n_\text{S}$, the lower flat band is empty, and the gap between this band and the next low energy band controls the transport. When the gaps close, a more complicated picture of partial occupation of multiple bands is needed. \\
The full e-h asymmetric full model leads to a partial filling of a band at $n=0$ for potential~$\Delta < \Delta_c$, i.e., the system is metallic in this regime. Beyond  $\Delta_c$, the gap keeps increasing with $\Delta$. This is seen in the experiments (see Fig. 2(d-f)) in the main manuscript).
On the other hand, e-h symmetric minimum model would lead to a CNP gap hence insulating behavior at arbitrarily small $\Delta$. This shows that one needs the full model to explain the experimental data.\\
This full model also captures the trend that the system remains insulating (i.e., finite \moire gap) at $n=\pm n_\text{S}$ up to a critical $\Delta$. The change in the gap as a function of $\Delta$ is not symmetric between $n=+n_\text{S}$ and $n=-n_\text{S}$, providing further evidence of e-h asymmetry in the system.\\ 
We note that our model overestimates the gap extracted by the experiments, but this is not surprising since transport gaps are often smaller than electronic structure gaps. However, the requirement of e-h asymmetry to explain the trends is a robust statement independent of the accuracy of the calculated gaps. 

\clearpage

\section{Extracting Arrhenius activation gap from temperature dependence of resistance at gaps}
\begin{figure}[hbt]
	\includegraphics[width=15cm]{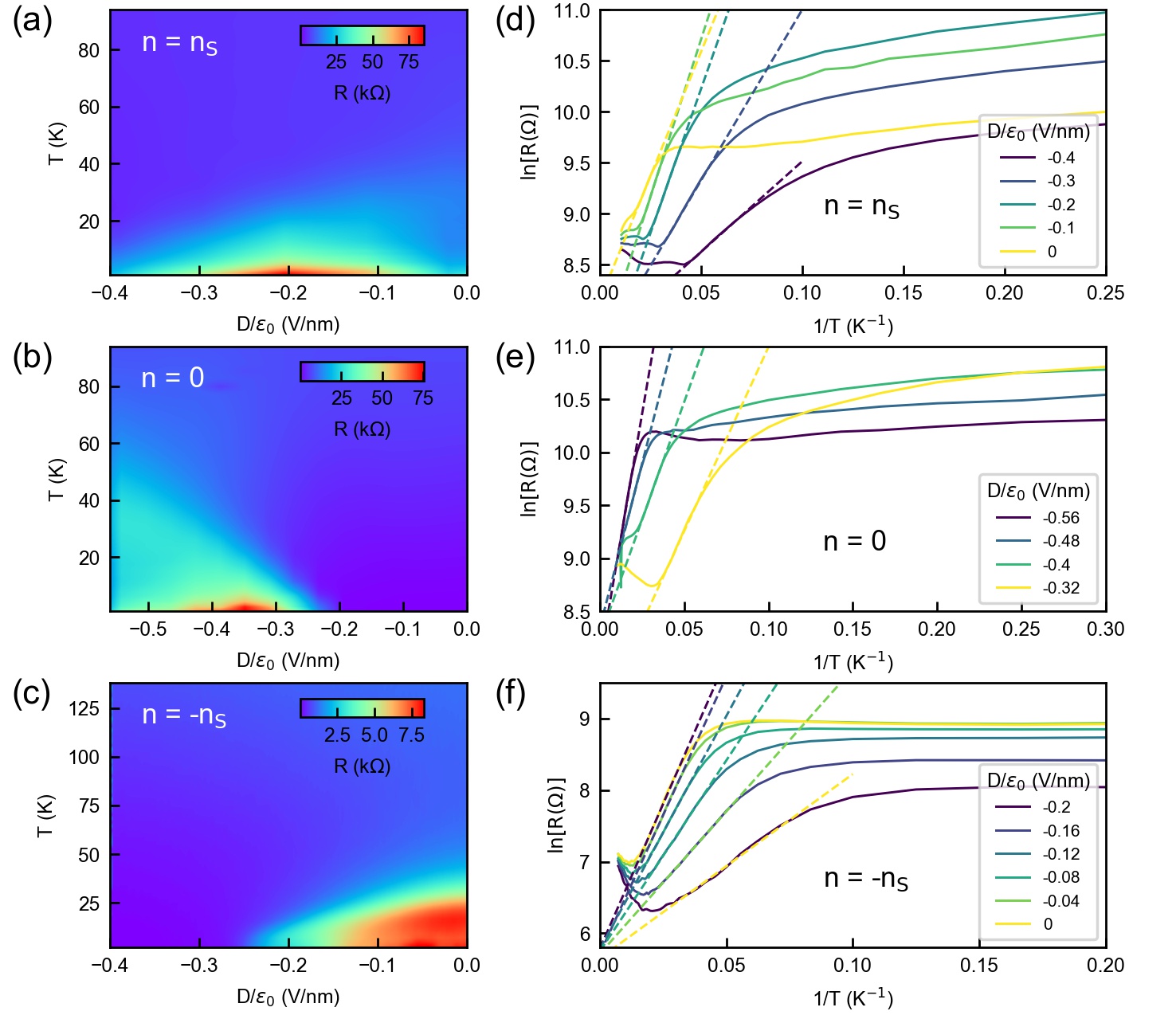}
	\caption{ \label{fig:fig_extended_data} {\footnotesize \textbf{Extraction of  Arrhenius gaps.} (a-c) Color scale plots of resistance as a function of temperature and electric displacement field for $n=n_{\rm S}$, $n=0$, and $n=-n_\text{S}$, respectively. (d-f) Extracting gap using Arrhenius equation, $R\sim \exp(\Delta/2 k_\text{B} T)$ by fitting the linear portion in $\ln{R}$ vs $1/T$ curves for $n=n_\text{S}$, $n=0$, and $n=-n_\text{S}$, respectively. The dashed lines are the fits.}}
\end{figure}

\clearpage

\section{Resistance variation at low temperature}
\begin{figure}[hbt]
	\includegraphics[width=15cm]{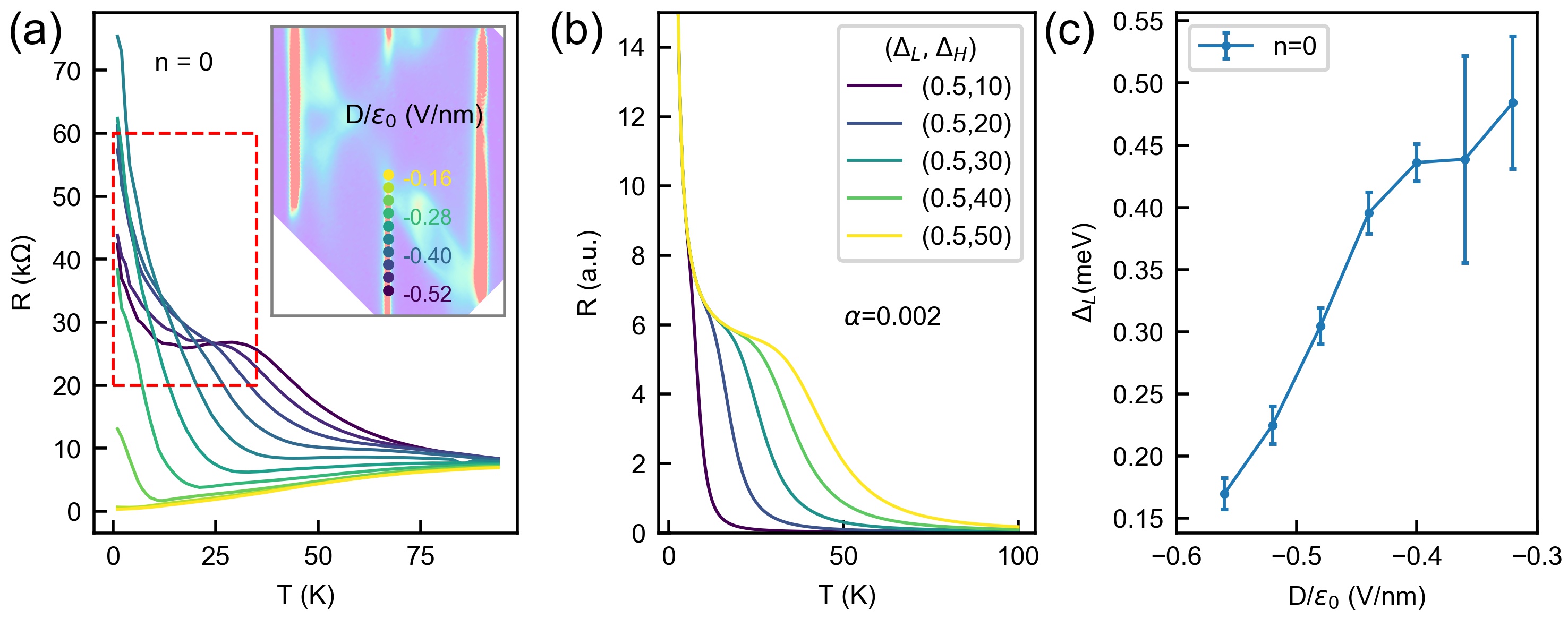}
	\caption{ \label{fig:fig_bump} {\footnotesize \textbf{Resistance variation at low temperature.} (a) Resistance vs. temperature at CNP for different electric displacement field. The red box indicates the regime where a bump appears in $R$ vs. $T$ curves. (b) Plots of simulated resistance for different combinations of $(\Delta_L, \Delta_H)$. (c) Extracted secondary gap, $\Delta_L$ by fitting Arrhenius equation to the temperature dependence of resistance at low temperature.  }}
\end{figure}

Here we focus our attention on the low-temperature regime of the temperature dependence of resistance with $n$ at gaps for various electric displacement fields, as shown in Fig. 2 in the main text. In particular, we would like to mention the observation of a bump in the temperature variation of resistance for $n=0$, as seen within the red square in Fig.~\ref{fig:fig_bump}(a). This feature is most prominent for high electric displacement field when the resistance varies non-monotonically with temperature showing a peak around $\sim 30$~K and a dip around $\sim 15$~K. While we can not rule out the possibility of low-temperature anomalies due to variable range hopping or existence of impurity levels, the correlation of this observation with electric displacement field is intriguing, especially since the electron-electron correlation induced gaps such as the gap at $n_\text{S}/2$ appears near the same electric displacement field. This feature can be explained within the semiconductor model~\cite{mermin} of transport where the conductance is given by, $\sigma(T)\sim(\alpha\exp(-\Delta_L/2k_\text{B}T)+\exp(-\Delta_H/2k_\text{B}T))$, which considers two channels of Arrhenius activation with different gaps and channel contributions. When $\Delta_L<<\Delta_H$ and $\alpha<<1$, one can see two activated behaviors, one at low $T$, and another at high  $T$, with a flat region in between. The functional dependence of $R$ vs. $T$ for this model is plotted in Fig.~\ref{fig:fig_bump}(b) for different combination of $(\Delta_L,\Delta_H)$ with $\alpha=0.002$. For our experimental data we find that this feature leads to the Arrhenius activation gap $\Delta_L \sim 0.5$ meV as plotted in Fig.~\ref{fig:fig_bump}(c).

\clearpage

\section{Data for another pair of probes}
\begin{figure}[hbt]
	\includegraphics[width=15cm]{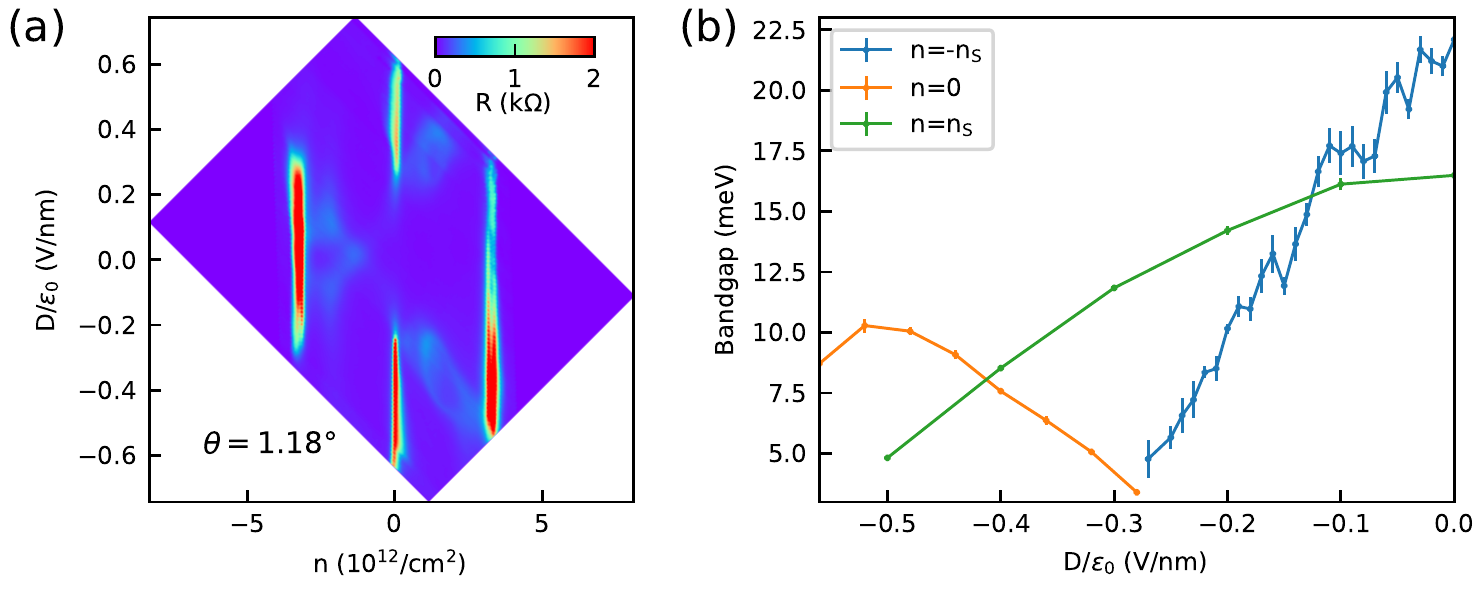}
	\caption{ \label{fig:fig_other_probe_data} {\footnotesize (a) Color scale plot of four-probe resistance as a function of carrier density and perpendicular electric displacement field. (b) Various gaps extracted from the temperature dependence of resistance at different electric displacement fields.}}
\end{figure}

\clearpage

\section{Extracting BANDWIDTH using the two band model} 
\begin{figure}[hbt]
	\includegraphics[width=15cm]{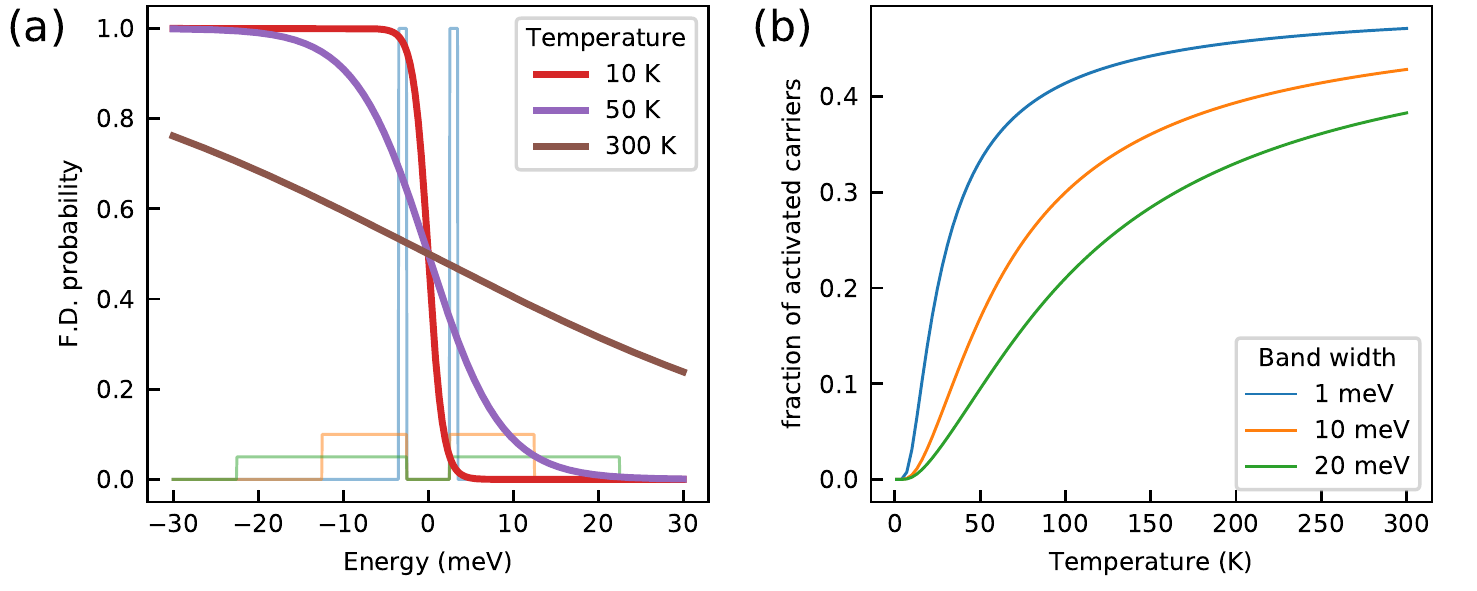}
	\caption{ \label{fig:fig_bandwidth} {\footnotesize (a) Fermi-Dirac probability of occupancy for three different temperatures with chemical potential at zero. Three sets of two-band DoS model is overlayed with bandwidths of 1 meV, 10 meV, and 20 meV to visualize the effect of flatness on saturation of activated charge carriers in a band. (b) The fraction of activated charge carriers against temperature for three different sets of flat bands.   }}
\end{figure}
As discussed in the main text, the low energy scales set by the narrow bandwidth of the flat bands and the small bandgap between them, the activated charge carriers  in the flat bands quickly saturate and the degradation of mobility with increasing temperature results in an insulator to metal crossover. While the detailed distribution of the density of states in a flat band is complicated, we associate a single energy scale, namely, the bandwidth of a rectangular shaped constant DoS flat band. To capture the actual density of states, one can associate more energy scales, but that will invoke more number of fitting parameters. 

To illustrate the interplay of the energy scales, we plot Fermi Dirac distribution probability in Fig.~\ref{fig:fig_bandwidth}(a) for three different temperatures. We overlay three sets of two-band DoS model  with bandwidths of 1 meV, 10 meV, and 20 meV and the same bandgap of 5 meV. For a band with narrower bandwidth, the DoS is distributed closer to the Fermi energy where the F.D probability quickly saturates to high-temperature value of 0.5. Effectively, the number of activated charge carriers in the narrower band saturates to its high-temperature value at a lower temperature. To visualize this effect more clearly, we plot the fraction of activated carries in a flat band with three different bandwidth in Fig.~\ref{fig:fig_bandwidth}(a). The fraction denotes the fraction of unoccupied states (holes) in the valence band or the occupied states (electrons) in the conduction band. This value saturates quickly to 0.5 for the narrower band.

\clearpage

\section{Metal to insulator crossover due to activation of carriers to HIGHER DISPERSING BANDS}
\begin{figure}[hbt]
	\includegraphics[width=15cm]{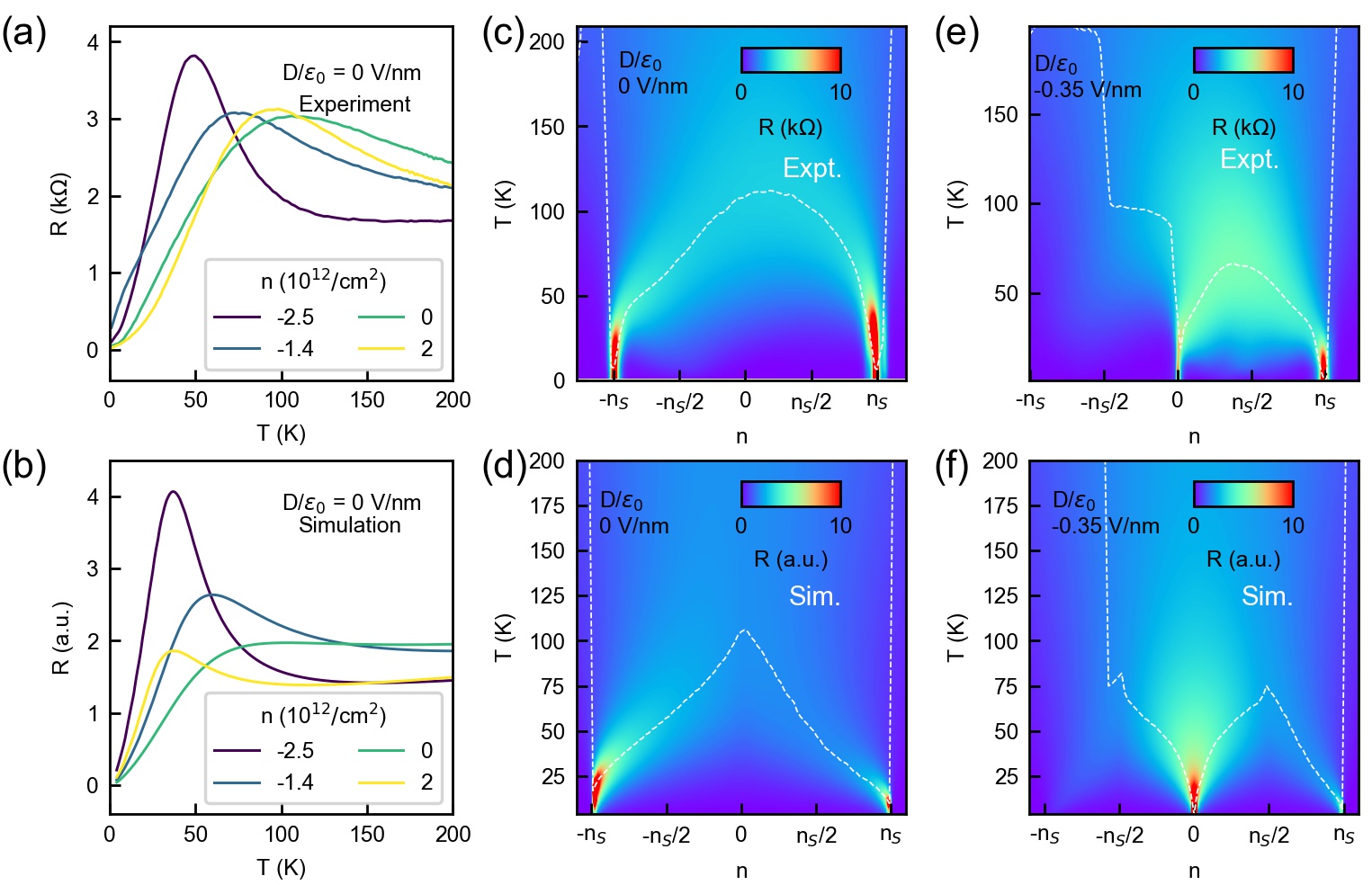}
	\caption{ \label{fig:fig_simulation} {\footnotesize \textbf{Simulating the effect of activation to higher band.} (a) Experimentally measured $R$ vs. $T$ for $D/\epsilon_0=0$ showing metal to insulator crossover for different $n$. (b) Simulated temperature dependence of resistance for different $n$ at $D/\epsilon_0=0$. (c) Color scale plots of experimentally measured resistance as a function of $n$ and $T$ for $D/\epsilon_0=0$. The overlay white dashed curve denotes the crossover temperature where resistance starts decreasing with increasing temperature. (d) Simulating metal to insulator crossover of (c). (e-f) Color scale plots similar to (c) and (d) respectively, corresponding to $D/\epsilon_0=-0.35$ V/nm.    }}
\end{figure}
As mentioned in the main text, and depicted in Fig.~\ref{fig:fig_simulation}(a), for the states not in the gaps, the temperature dependence of resistance shows metallic behavior initially and starts showing insulating behavior after a crossover temperature. This metal to insulator transition can be understood as an effect of activation of charge carriers to dispersing bands from the flat bands. Since the mobility of the dispersing bands is greater than the flat bands, the activated charge carriers start contributing  dominantly in the transport and as the temperature is further increased, the number of activated carriers increases showing insulating behavior. We simulate this crossover by calculating contribution in conductance  from different bands by using the simple formula $\sigma=n e \mu$ with $e$ being the elementary  charge. Here, $\mu=a T^{-3/2}$ is the mobility with $a$ being a constant which is lower for the flat band. The number of carriers is calculated by $n= \int\text{DoS}(E) F(E,T) dE$, where $F (E,T)$ is the Fermi-Dirac probability of electrons (holes) in the conduction (valence) band. For DoS, we use constant density of states in different bands. We take the values of the gaps as extracted from the measurement. In Fig.~\ref{fig:fig_simulation}(b), we plot  resistance against temperature simulated for different $n$ at $D/\epsilon_0=0$.  These curves follow a similar trend as the experimentally measured curves shown in Fig.~\ref{fig:fig_simulation}(a). In Fig.~\ref{fig:fig_simulation}(c) and (e) we show the color-scale plots of experimentally measured resistance as a function of temperature and charge density for two different electric displacement fields, $D/\epsilon_0=0$  and  $D/\epsilon_0=-0.35$ V/nm respectively. In Fig.~\ref{fig:fig_simulation}(d) and (f) we show the simulated data corresponding to Fig.~\ref{fig:fig_simulation}(c) and (e).

\clearpage

\section{Temperature dependence data for TDBG DEVICE WITH A TWIST ANGLE OF 2.05$^{\circ}$}
\begin{figure}[hbt]
	\includegraphics[width=15cm]{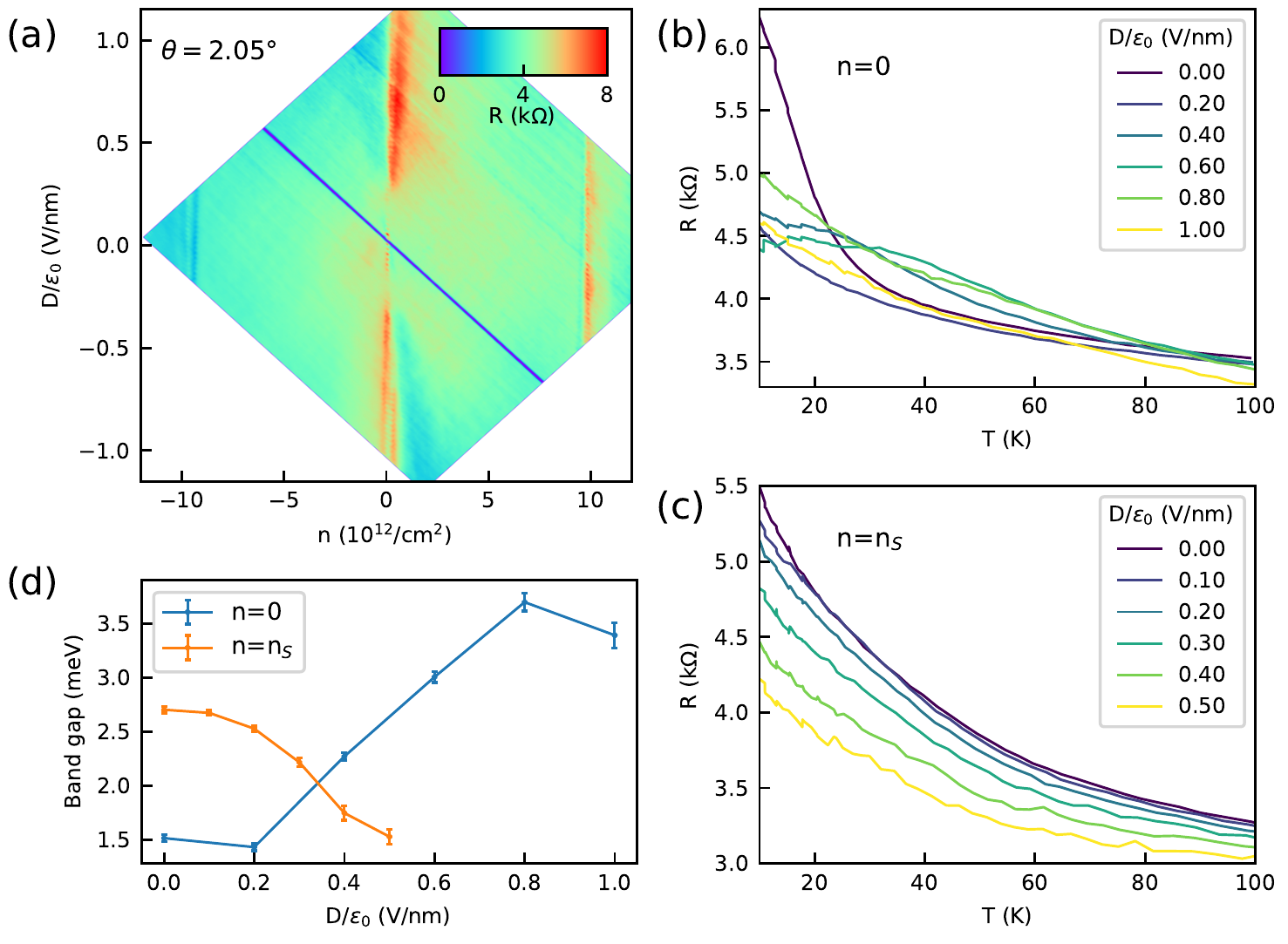}
	\caption{ \label{fig:fig_two_degree_dev} {\footnotesize \textbf{Data for a TDBG device with twist angle of 2.05$^{\circ}$.} (a) Color scale plot of four-probe resistance as a function of carrier density, $n$ and perpendicular electric displacement field, $D$ at 1.5 K. (b) Variation of resistance as a function of temperature at the charge neutrality point, $n=0$ for different electric displacement fields. (c) Resistance vs. temperature for different electric displacement fields at $n=n_\text{S}$. (d) Extracted gaps by fitting $R$ vs. $T$ to Arrhenius activation equation $R\sim \exp(\Delta/2 k_\text{B} T)$. }}
\end{figure}
In Fig.~\ref{fig:fig_two_degree_dev} we present data for a TDBG device with a twist angle of 2.05$^{\circ}$ corresponding to $n_\text{S} = 9.88 \times 10^{12}$ cm$^{-2}$. The variation of resistance at 1.5 K with charge density and electric displacement field, as plotted in Fig.~\ref{fig:fig_two_degree_dev}(a), shows the single-particle energy gaps, namely, the two \moire gaps at $n=\pm n_\text{S}$ and CNP gap at $n=0$. However, there is no correlation induced gap seen in the parameter space we have probed. Unlike the TDBG device with a twist angle of 1.18$^{\circ}$ , this device shows resistance peak at CNP even at zero electric displacement field, indicating the existence of a bandgap. This gap closes and opens up again upon further increase of the perpendicular electric displacement field.  Similar trend has been observed in a recent study~\cite{asvinspin}. Figures~\ref{fig:fig_two_degree_dev}(b) and (c) show the variation of resistance with temperature for  $n=0$ and $n=n_\text{S}$ respectively, showing  insulating behavior throughout the temperature regime we explored. In Fig.~\ref{fig:fig_two_degree_dev}(d), we plot the corresponding gaps extracted by fitting Arrhenius activation.

\clearpage

\section{Electric field dependence of CNP gap for bilayer graphene vs. TDBG system}
\begin{figure}[hbt]
	\includegraphics[width=8cm]{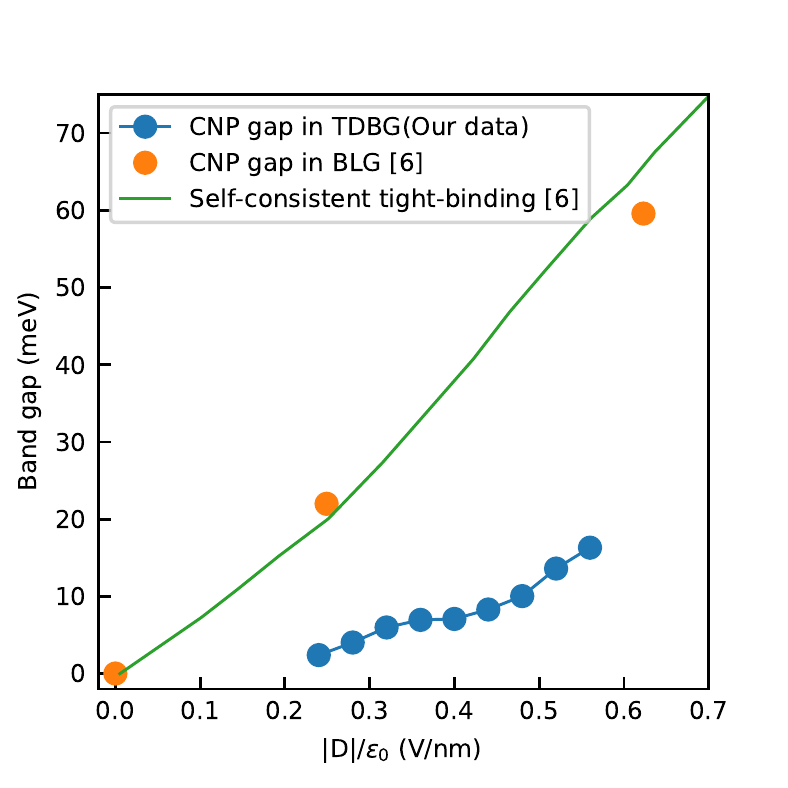}
	\caption{ \label{fig:gapcompare} {\footnotesize A comparison of the trends for evolution of gap at the charge neutrality point with electric filed between two different systems -- twisted double bilayer graphene (TDBG) with twist angle $1.18\degree$ as we measure in this work and bilayer graphene (BLG) from Y. Zhang et al.\cite{zhang_direct_2009}.}}
\end{figure}
\clearpage